\documentclass[aps,prl,reprint,twocolumn,showpacs,superscriptaddress]{revtex4}

\usepackage[dvips]{graphicx}
\usepackage{dcolumn}
\usepackage{bm}
\usepackage{xspace}
\usepackage{multirow}
\usepackage{natbib}
\usepackage{color}

\begin{document}

\preprint{}
\title{Electronic structure of superconducting KC$_8$
           and non-superconducting LiC$_6$ graphite intercalation
           compounds: Evidence for a graphene-sheet-driven
           superconducting state}

\author{Z.-H. Pan}
\author{J. Camacho}
\affiliation{Condensed Matter Physics and Materials Science Department, Brookhaven National Lab, Upton, NY 11973}
\author{M.H. Upton}
\affiliation{Advanced Photon Source, Argonne National Laboratory, Argonne, IL 60192}
\author{A. V. Fedorov}
\affiliation{Advanced Light Source, Lawrence Berkeley National Laboratory, Berkeley, CA 94720}
\author{C.A. Howard}
\author{M. Ellerby}
\affiliation{London Centre for Nanotechnology and Department of Physics and Astronomy, University College London, London WC1E 6BT, United Kingdom}
\author{T. Valla}
\email{valla@bnl.gov}
\affiliation{Condensed Matter Physics and Materials Science Department, Brookhaven National Lab, Upton, NY 11973}

\date{\today}

\begin{abstract}
We have performed photoemission studies of the electronic structure in LiC$_6$ and KC$_8$, a non-superconducting and a superconducting graphite intercalation compound, respectively. We have found that the charge transfer from the intercalant layers to graphene layers is larger in KC$_8$ than in LiC$_6$, opposite of what might be expected from their chemical composition. We have also measured the strength of the electron-phonon interaction on the graphene-derived Fermi surface to carbon derived phonons in both materials and found that it follows a universal trend where the coupling strength and superconductivity monotonically increase with the filling of graphene $\pi^{\ast}$ states. This correlation suggests that both graphene-derived electrons and graphene-derived
phonons are crucial for superconductivity in graphite intercalation compounds.
\end{abstract}
\vspace{1.0cm}


\maketitle
\pagebreak

In graphite intercalation compounds (GIC), the intercalation of various atomic 
or molecular species in between graphene layers in graphite leads to novel 
properties and a very rich physics, including superconductivity 
\cite{dresselhaus2002}. In graphite intercalated with alkaline metals, 
superconductivity has been known for decades \cite{hannay1965}, but after recent 
discovery of relatively high $T_c$ superconductivity in CaC$_6$ ($T_c=11.5$ 
K) \cite{weller2005,emery05} research in this field has been intensified.
Even though the electron-phonon coupling (EPC) is most likely responsible for 
pairing in GICs \cite{kim2006,lamura2006,hinks2007}, it is still not clear what 
electronic states, intercalant- or graphene- derived ones, and what phonons are 
responsible for pairing \cite{mazin2005,calandra2005,boeri2007,mazin2007}. Due 
to differences in structure and composition, no clear trends have been 
identified that could unambiguously resolve these issues. For example, KC$_8$ is 
a superconductor and LiC$_6$ is not. Further, in GICs intercalated with alkaline 
earths, $T_c$ ranges from zero to 11.5K, even though they share the same 
chemical formula MC$_6$, where M is an alkaline earth atom. 
On the other hand, band structure calculations show that in graphite and GICs, 
an interlayer state exists above $\pi^{\ast}$ band 
\cite{posternak1983,holzwarth1984}, prompting some researchers to propose that 
its partial filling and coupling to soft intercalant phonons induces 
superconductivity in GICs \cite{csanyi2005,mazin2005}.
The experimental situation is still inconclusive, with strong advocates for 
intercalant \cite{hinks2007} and graphene dominated superconductivity 
\cite{kim2006,valla2009,gruneis2009,dean2010}. 
Recent angle resolved photoemission spectroscopy (ARPES) study on CaC$_6$ 
\cite{valla2009} reported that EPC on graphene-derived Fermi surface (FS) to 
graphene phonons is strong enough to explain a $T_c$ in the range of tens of 
Kelvin, indicating that graphene sheets provide crucial ingredients for 
superconductivity in GICs.
However, to test this idea, it would be important to extend similar studies to 
GICs with different $T_c$.

In this letter, we report ARPES studies of the electronic structure and the EPC 
in the non-superconducting LiC$_6$ and in superconducting KC$_8$ ($T_c=0.39$ K) 
and compare these materials with several other GICs. We find that the EPC on the 
graphene $\pi^{\ast}$ states to the graphene derived phonons increases 
with the filling of $\pi^{\ast}$ states in a sequence from LiC$_6$ to KC$_8$ to 
CaC$_6$, following the same trend as $T_c$. The positive correlation between 
these quantities implies that superconductivity originates in graphene sheets 
while the main role of intercalants is to provide the charge for the 
graphene states. 

The experiments were carried out on a Scienta SES-100 electron spectrometer 
operating in the angle-resolved mode at the beamline 12.0.1 of the Advanced 
Light Source. The spectra were recorded at the photon energy of 50 eV, 
with the combined instrumental energy resolution of ~ 20-25 meV and 
the momentum resolution of $\pm 0.008$ \AA$^{-1}$ in geometry where the 
polarization of light was perpendicular to the probed momentum line. The LiC$_6$ 
and KC$_8$ samples were prepared by intercalating natural, single-crystal 
graphite flakes (Madagascan) using the vapour transport method as described in 
Ref. \cite{dresselhaus2002}. X-ray diffraction showed very 
high sample purity with no graphite or secondary stage phases. 
To avoid degradation, all samples were unsealed 
and glued to the sample holder with Ag-epoxy in an Ar filled glow box. Protected 
by the epoxy, they were then quickly transferred to the ARPES prep-chamber, 
and cleaved at low temperature (15-20 K) under ultra-high vacuum conditions 
($2\times10^{-9}$ Pa). All data were collected at 15-20 K. 

\begin{figure}[htbp]
\begin{center}
\includegraphics[width=7.5cm]{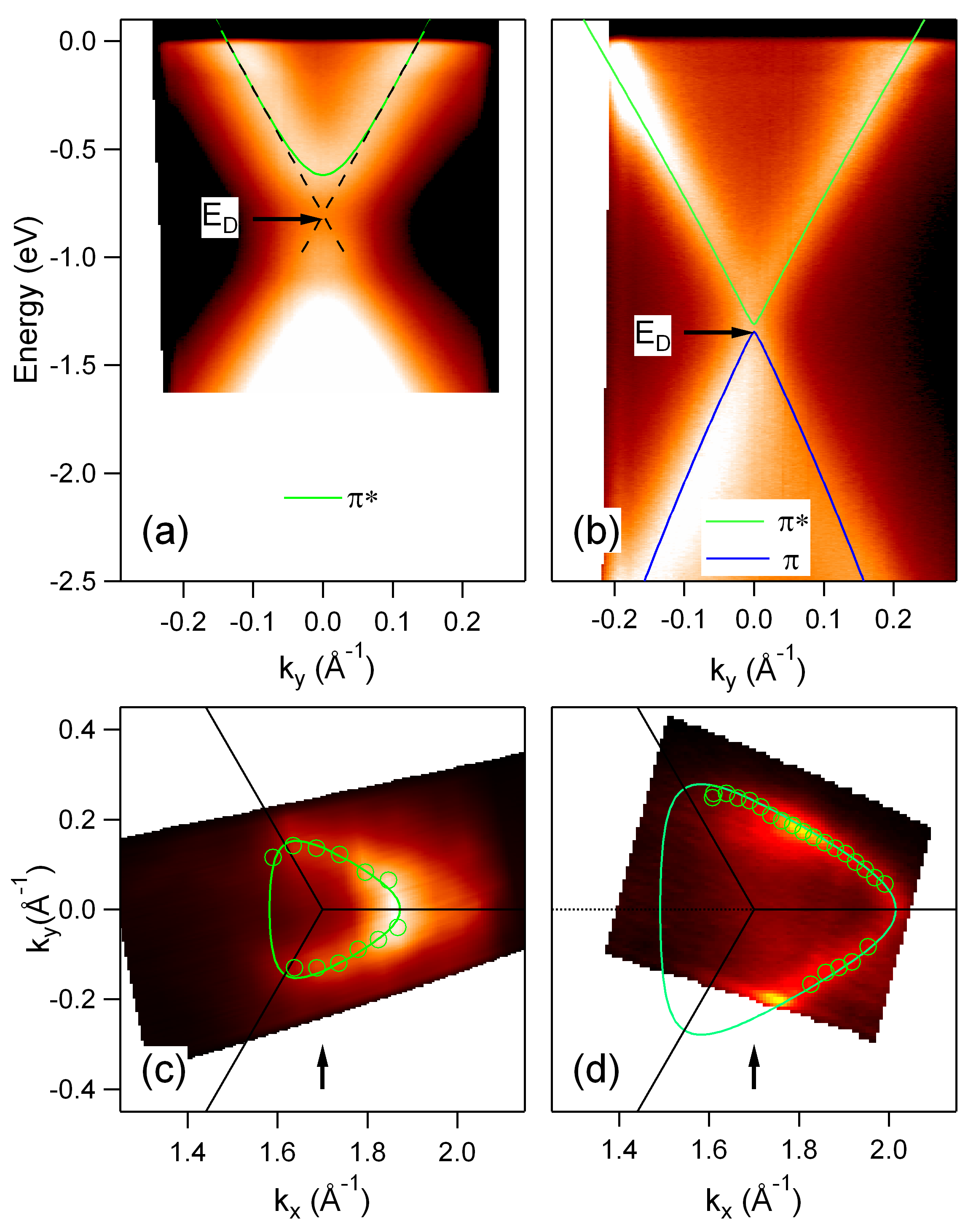}
\caption{Photoemission spectra from LiC$_6$ (a) and KC$_8$ (b) along the same 
momentum line in the graphene Brilluoin Zone traversing the K point, as 
indicated by the arrows in panels (c) and (d). Lines represent the $\pi$ and 
$\pi^{\ast}$ bands. Arrows indicate the binding energy of Dirac point. (c) and 
(d) Photoemission intensity from a narrow energy interval around the Fermi level 
($\omega=\pm 10$ meV), representing the graphene-derived $\pi^{\ast}$  FS, for 
LiC$_6$ and KC$_8$, respectively. Circles represent the MDC peak positions, 
while lines represent the tight-binding fits to the data, as described in the 
text. 
}
\label{Fig1}
\end{center}
\end{figure}

Fig.~\ref{Fig1} shows the ARPES spectra near the K point in the graphene 
Brillouin zone (BZ) for LiC$_6$  and KC$_8$. The upper panels (a) and (b) show 
the contours of photoemission intensity as a function of binding energy for a 
momentum line going through the K point. The intensity from a narrow interval 
($\pm 10$ meV) around the Fermi level, representing the FS, is shown in the 
lower panels (c) and (d).  
The dispersing states are the graphene-derived $\pi$ and $\pi^{\ast}$ bands, as 
marked in Fig.~\ref{Fig1}(a) and (b). In KC$_8$ the low energy band structure is 
essentially graphene-like, with $\pi$ and $\pi^{\ast}$ bands touching at the 
Dirac point \cite{gruneis2009} which is shifted below the Fermi level due to 
doping. 
In LiC$_6$, a sizable gap exists between $\pi$ and $\pi^{\ast}$ band.
Dirac point is determined by extrapolating the linear part of $\pi^{\ast}$ 
dispersion at low binding energies to the K point. 
The arrows in Fig.~\ref{Fig1}(a) and (b) indicate the position of Dirac point, 
(E$_D$), 0.825 eV and 1.35 eV for LiC$_6$ and KC$_8$, respectively. 
It is clear from Fig.~\ref{Fig1} that the $\pi^{\ast}$ band is filled more in 
KC$_8$ than in LiC$_6$  and that it forms a larger FS in the former material. 
The area enclosed by the FS (lower panels in Fig.~\ref{Fig1}) is a direct 
measure of doping of graphene $\pi^{\ast}$ states, i.e. of the charge 
transferred into the graphene sheets. The FS is determined from peak positions 
of momentum distribution curves (MDCs) at $E_F$ (open circles) and compared to 
the $3^{rd}$ nearest neighbor hopping tight binding band structure (lines). In 
KC$_8$, the occupied FS area is 0.399 \AA$^{-2}$ which corresponds to 0.11 
electrons per graphene unit cell (GUC), or 44\% of the nominal value of 0.25 for 
the complete charge transfer. In LiC$_6$, the occupied $\pi^{\ast}$ (green 
circles and lines) FS area is $0.125$ \AA$^{-2}$, corresponding to the doping of 
only 0.0344 electrons per GUC. This is far below the nominal value of 0.33 
electrons per GUC. 

The incomplete charge transfer into the graphene $\pi^{\ast}$ states would 
suggest that the remaining charge occupies the so-called interlayer band. The 
occupation of the interlayer state has been recently reported in 
CaC$_6$ \cite{sugawara2009}, but the observed feature was very weak and broad. 
Even the graphene-derived $\pi^{\ast}$ state was very broad and did not form an 
enclosed contour at the Fermi level, casting doubts on these results.
Our experiments always show relatively sharp $\pi^{\ast}$ band that forms a well 
defined FS. However, in MC$_6$(M=Li, Ca, Ba) materials, in addition to 
$\pi^{\ast}$ band, we always see a broad feature at slightly higher binding 
energy that follows the $\pi^{\ast}$ band, dispersing upward from K point 
(Fig.~\ref{Fig1}(a)). Further from the K point, the broad feature loses 
intensity and its dispersion cannot be precisely traced. In CaC$_6$, this 
feature is observable over larger region of $k$ space (Fig. 1 in 
\cite{valla2009}). 
It is possible be that this feature is a remnant of an interlayer band, smeared 
out by a disorder within the intercalant layers and folded into the K point of 
the graphene BZ. However, our measurements do not show any evidence of the 
interlayer band in the region from which it should be folded to the K point - 
the $\Gamma$ point. 
We note that in pristine graphite, the interlayer hopping $t_{\perp}$ splits 
both $\pi$ and $\pi^{\ast}$ bands into the bonding and antibonding counterparts 
due to the $AB$ stacking of graphene sheets. It would be tempting to assign the 
broad feature as a bonding $\pi^{\ast}$ state, due to similarities in initial 
dispersion. However, all first stage GICs have the $AA$ stacking of graphene 
sheets and such assignment would be incorrect.
On the other hand, in LiC$_6$ and CaC$_6$ the $\pi^{\ast}$ 
band might be split due to the AB stacking of the intercalant (if there is not 
too much disorder in the intercalant sites). If this was indeed the case, the correct charge 
transfer to the graphene layers would be 0.0616 and 0.349 electrons per GUC, 
for LiC$_6$ and for CaC$_6$, respectively. 

Irrespective of these issues, our experimental observation that the doping of 
graphene sheets is larger in KC$_8$ than in LiC$_6$ is opposite of the expected 
nominal doping, but is in line with the existence of superconductivity in these 
materials: KC$_8$ is a superconductor and LiC$_6$ is not. 
In the following, we identify the reason for the correlation between 
superconductivity and doping of the graphene sheets.
It is evident from Fig. 1 that in both LiC$_6$ and KC$_8$, an anomaly or a kink 
in dispersion of the $\pi^{\ast}$ band occurs at approximately 0.165 eV below 
the $E_F$. This is a hallmark of the interaction of the electronic states with 
phonons \cite{valla1999} that have been previously observed in CaC$_6$ and 
KC$_8$ \cite{valla2009,gruneis2009} and attributed to a coupling to graphene in-
plane high-frequency phonons. 
To quantify the electron phonon coupling, we have used the standard MDC fitting 
procedure \cite{lashell2000,valla2000} which uses a tight binding dispersion as 
the starting approximation for the bare band and gives the real 
($\text{Re}\Sigma$) and imaginary ($\text{Im}\Sigma$) part of self energy as 
fitting parameters. The bare band dispersion is then refined until the obtained 
$\text{Re}\Sigma$ and $\text{Im}\Sigma$ satisfy Kramers-Kronig transformations 
\cite{kordyuk2005}. 
Panels (a) and (c) in Fig. 2 show the $\text{Im}\Sigma$ while panels (b) and (d) 
show $\text{Re}\Sigma$ for both materials for several different locations on the 
FS, as indicated in figure.

\begin{figure}[h]
\begin{center}
\includegraphics[width=7cm]{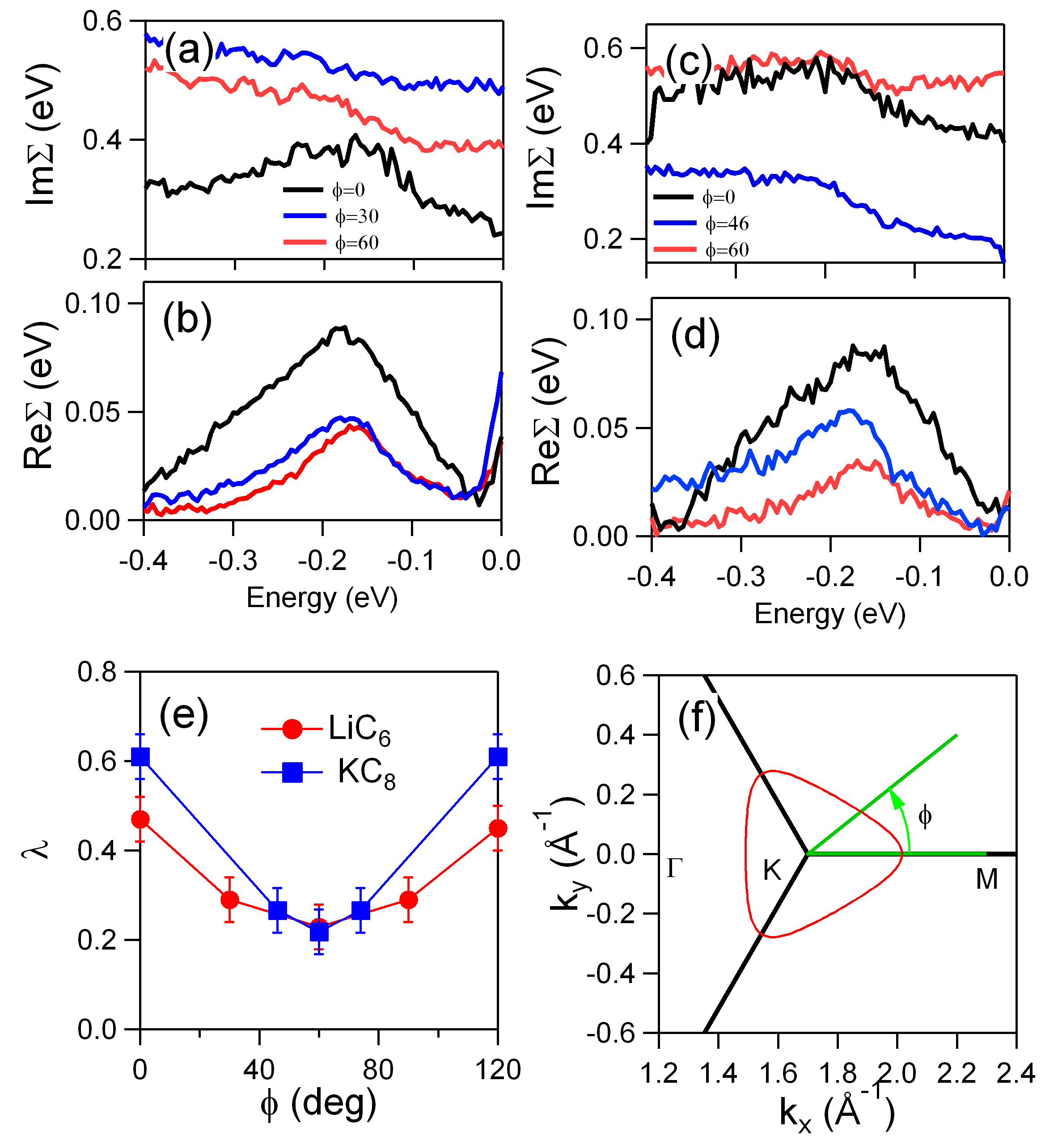}
\caption{Im$\Sigma(\omega)$ (a) and Re$\Sigma(\omega)$ (b) for LiC$_6$ and KC$_8$ ((c) 
and (d)) for several different points at the FS, as indicated in (e) and (f). 
(e) the electron-phonon coupling strength, $\lambda$, for LiC$_6$ and KC$_8$, 
extracted from Re$\Sigma(\omega)$ as a function of polar angle $\phi$ as defined 
in (f).}
\label{Fig2}
\end{center}
\end{figure}

$\text{Re}\Sigma$ in both materials shows a peak at around -0.165 eV, while 
$\text{Im}\Sigma$ shows a decrease below that energy, indicating a coupling to 
the phonon mode. The only phonons with such high energy are graphene-derived in-plane
 phonon modes. A small variation in the energy at which $\text{Re}\Sigma$ 
has a maximum at different points on the FS indicates a slight dispersion of the 
mode. We note that the sharp increase of Re$\Sigma$ below 20 meV is 
an artifact of finite energy resolution of the experimental apparatus \cite{valla2006}. We have excluded the affected interval $|\omega|<$ 20 meV from the considerations and any fine structure, related to a possible coupling to the intercalant modes, is out of our detection limits. However, the lack of broadening of the $\pi^{\ast}$ states with increasing temperature over the range of 15 K $<$ T $<$ 200 K suggests that the low energy modes play insignificant role and that the EPC is dominated by the graphene-derived high frequency modes. 
\begin{figure}[h]
\begin{center}
\includegraphics[width=5.2cm]{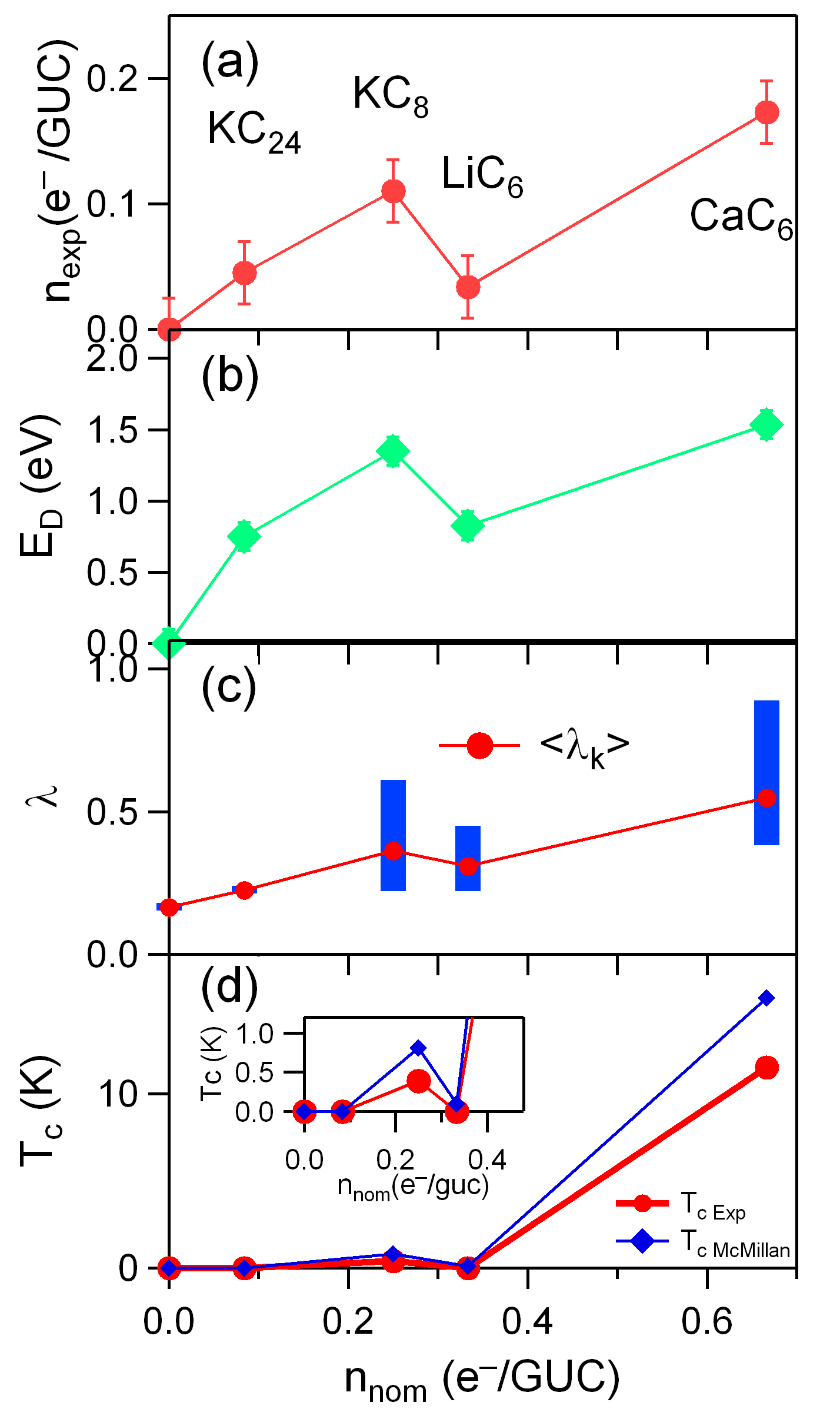}
\caption{(a) Measured charge transfer (electrons per graphene unit cell (GUC)) from the 
intercalant to the graphene $\pi^{\ast}$ states. (b) Binding energy of Dirac 
point. (c) EPC coupling constant $\lambda$. The range 
of each vertical bar indicates the anisotropy with the maximum along the KM line 
and minimum along the K$\Gamma$ line.
Red circles represent the momentum-averaged coupling constant $\langle\lambda_{k}\rangle$. 
(d) Superconducting transition temperature $T_c$ in GICs. Red dots represent experimental values \cite{weller2005,emery05,hannay1965}, and blue diamonds are calculated from the measured $\langle\lambda_{k}\rangle$, shown in (c), using McMillan's formula \cite{McMillan1968}.
The inset zooms in the low $T_c$ values. All the quantities are shown as functions of nominal chemical 
composition for measured GICs. Zero corresponds to the pristine graphite. Data 
for CaC$_6$ and pristine graphite are from ref. \cite{valla2009} and for 
KC$_{24}$ are from ref. \cite{camacho}.}
\label{Fig3}
\end{center}
\end{figure}

The coupling constant $\lambda$ can be extracted directly from 
$\text{Re}\Sigma$ as ($\lambda=-
[\partial(\text{Re}\Sigma)/\partial\omega]_0$) by fitting the low energy part to 
a straight line. It shows some 
anisotropy [Fig.~\ref{Fig2}(e)], with the maximum along the KM direction and the 
minimum along the $\Gamma$K direction, similar, but significantly smaller than 
in CaC$_6$ \cite{valla2009} and what was recently reported for KC$_8$ 
\cite{gruneis2009}. The most important observation, however, is that the 
momentum averaged $\langle\lambda_{k}\rangle$ is stronger in KC$_8$ than in 
LiC$_6$. The coupling constant and its anisotropy both increase from LiC$_6$ to 
KC$_8$ to CaC$_6$, exactly in the same sequence as the filling of the graphene 
$\pi^{\ast}$ band and in previously established sequence for $T_c$. 
Strengthening of EPC with the filling of the $\pi^{\ast}$ band has also been 
observed in the epitaxial graphene \cite{mcchesney2007}. 
This is not surprising because the density of states near the $E_F$ increases 
with the filling of the $\pi^{\ast}$ band and a larger FS makes an EPC process 
more probable as the phase space available for the scattering events grows. In 
the pristine graphite, the FS is nearly a point and the EPC is strongly 
suppressed \cite{valla2009,leem2008}.

To better illustrate a positive correlation between $T_c$, $\lambda$ and doping 
of the graphene $\pi^{\ast}$ band in different GICs, we plot these quantities in 
Fig.~\ref{Fig3} as functions of nominal chemical composition for several 
different materials.
Actual (measured) charge transfer (in electrons/GUC) is shown in 
Fig.~\ref{Fig3}(a). 
The increase in size of the FS is consistent with the energy of Dirac point 
[panel (b)] as the chemical potential, $\mu$, shifts from the pristine graphite 
to CaC$_6$. It is interesting to note that $\mu\propto\sqrt n_{exp}$ still holds, 
regardless of the shape of FS and number of FS sheets in these five different 
materials. 
The coupling constant $\lambda$ and $T_c$ follow the 
same trend. This suggests that the graphene $\pi^{\ast}$ states and their coupling to graphene in-plane phonons is crucial for superconductivity and that the only role that the intercalants seem to play is to provide the charge for filling of the $\pi^{\ast}$ states. 
This is further re-inforced by the calculated $T_c$ using McMillan's formula \cite{McMillan1968}:
\begin{equation}
T_c=\frac{\Theta}{1.45}\textrm{exp}(-\frac{1.04(\lambda+1)}{\lambda-(0.62\lambda+1)\mu^{\ast}})
\end{equation}
where we use measured $\langle\lambda_{k}\rangle$ (panel(c)), Debye teperature $\Theta=$1926 K and Coulomb pseudo-potential $\mu^{\ast}\sim 0.14$. As shown in Fig.~\ref{Fig3}(d), the calculated $T_c$ values are very close to the ones measured experimentally. 
The threshold-like behavior of $T_c$ near $\langle\lambda_k\rangle=0.3$ places LiC$_6$ on one side and KC$_8$ on 
another side of a steep increase in $T_c$.
We note that superconductivity in LiC$_3$ and LiC$_2$, materials in which more 
Li is pushed in under pressure, supports our picture where the EPC and 
superconductivity strengthen with the filling of graphene $\pi^{\ast}$ states. 
The increase in $T_c$ from 0.39 K for stoichiometric KC$_8$ to 0.55K in material 
with excess K is also in line with this picture. A further test would be a 
systematic ARPES study on alkaline-earth GICs (Ca, Sr, Ba) where $T_c$ decreases 
with the atomic mass of alkaline-earth intercalant. 

In conclusion, we have identified the universal trend in alkali and alkaline-
earth GICs where superconductivity is tightly correlated with the doping of 
graphene-derived $\pi^{\ast}$ states and with the coupling of these states to 
graphene phonons. This implies that the graphene sheets play the crucial 
role in superconductivity in GICs. 

We acknowledge useful discussions with M. Calandra, M. Dean, M. Khodas, E. Rotenberg, M. Strongin 
and A. Walters. Work at Brookhaven is
supported by the US DOE under Contract No. DE-AC02-98CH10886. 
Work at University College London is supported by the UK Engineering and Physical 
Science Research Council. ALS is operated by the US DOE under
Contract No. DE-AC03-76SF00098.


\providecommand{\noopsort}[1]{}\providecommand{\singleletter}[1]{#1}%
\begin{thebibliography}{28}
\expandafter\ifx\csname natexlab\endcsname\relax\def\natexlab#1{#1}\fi
\expandafter\ifx\csname bibnamefont\endcsname\relax
  \def\bibnamefont#1{#1}\fi
\expandafter\ifx\csname bibfnamefont\endcsname\relax
  \def\bibfnamefont#1{#1}\fi
\expandafter\ifx\csname citenamefont\endcsname\relax
  \def\citenamefont#1{#1}\fi
\expandafter\ifx\csname url\endcsname\relax
  \def\url#1{\texttt{#1}}\fi
\expandafter\ifx\csname urlprefix\endcsname\relax\def\urlprefix{URL }\fi
\providecommand{\bibinfo}[2]{#2}
\providecommand{\eprint}[2][]{\url{#2}}

\bibitem[{\citenamefont{Dresselhaus and Dresselhaus}(2002)}]{dresselhaus2002}
\bibinfo{author}{\bibfnamefont{M.}~\bibnamefont{Dresselhaus}} \bibnamefont{and}
  \bibinfo{author}{\bibfnamefont{G.}~\bibnamefont{Dresselhaus}},
  \bibinfo{journal}{Adv. Phys.} \textbf{\bibinfo{volume}{51}},
  \bibinfo{pages}{1} (\bibinfo{year}{2002}).

\bibitem[{\citenamefont{Hannay et~al.}(1965)\citenamefont{Hannay, Geballe,
  Matthias, Andres, Schmidt, and MacNair}}]{hannay1965}
\bibinfo{author}{\bibfnamefont{N.~B.} \bibnamefont{Hannay}},
  \bibinfo{author}{\bibfnamefont{T.~H.} \bibnamefont{Geballe}},
  \bibinfo{author}{\bibfnamefont{B.~T.} \bibnamefont{Matthias}},
  \bibinfo{author}{\bibfnamefont{K.}~\bibnamefont{Andres}},
  \bibinfo{author}{\bibfnamefont{P.}~\bibnamefont{Schmidt}}, \bibnamefont{and}
  \bibinfo{author}{\bibfnamefont{D.}~\bibnamefont{MacNair}},
  \bibinfo{journal}{Phys. Rev. Lett.} \textbf{\bibinfo{volume}{14}},
  \bibinfo{pages}{225} (\bibinfo{year}{1965}).

\bibitem[{\citenamefont{Weller et~al.}(2005)\citenamefont{Weller, Ellerby,
  Saxena, Smith, and Skipper}}]{weller2005}
\bibinfo{author}{\bibfnamefont{T.}~\bibnamefont{Weller}},
  \bibinfo{author}{\bibfnamefont{M.}~\bibnamefont{Ellerby}},
  \bibinfo{author}{\bibfnamefont{S.}~\bibnamefont{Saxena}},
  \bibinfo{author}{\bibfnamefont{R.}~\bibnamefont{Smith}}, \bibnamefont{and}
  \bibinfo{author}{\bibfnamefont{N.}~\bibnamefont{Skipper}},
  \bibinfo{journal}{Nat. Phys.} \textbf{\bibinfo{volume}{1}},
  \bibinfo{pages}{39} (\bibinfo{year}{2005}).

\bibitem[{\citenamefont{Emery et~al.}(2005)\citenamefont{Emery, H\'erold,
  d'Astuto, Garcia, Bellin, Mar\^ech\'e, Lagrange, and Loupias}}]{emery05}
\bibinfo{author}{\bibfnamefont{N.}~\bibnamefont{Emery}},
  \bibinfo{author}{\bibfnamefont{C.}~\bibnamefont{H\'erold}},
  \bibinfo{author}{\bibfnamefont{M.}~\bibnamefont{d'Astuto}},
  \bibinfo{author}{\bibfnamefont{V.}~\bibnamefont{Garcia}},
  \bibinfo{author}{\bibfnamefont{C.}~\bibnamefont{Bellin}},
  \bibinfo{author}{\bibfnamefont{J.~F.} \bibnamefont{Mar\^ech\'e}},
  \bibinfo{author}{\bibfnamefont{P.}~\bibnamefont{Lagrange}}, \bibnamefont{and}
  \bibinfo{author}{\bibfnamefont{G.}~\bibnamefont{Loupias}},
  \bibinfo{journal}{Phys. Rev. Lett.} \textbf{\bibinfo{volume}{95}},
  \bibinfo{pages}{087003} (\bibinfo{year}{2005}).

\bibitem[{\citenamefont{Kim et~al.}(2006)\citenamefont{Kim, Kremer, Boeri, and
  Razavi}}]{kim2006}
\bibinfo{author}{\bibfnamefont{J.~S.} \bibnamefont{Kim}},
  \bibinfo{author}{\bibfnamefont{R.~K.} \bibnamefont{Kremer}},
  \bibinfo{author}{\bibfnamefont{L.}~\bibnamefont{Boeri}}, \bibnamefont{and}
  \bibinfo{author}{\bibfnamefont{F.~S.} \bibnamefont{Razavi}},
  \bibinfo{journal}{Phys. Rev. Lett.} \textbf{\bibinfo{volume}{96}},
  \bibinfo{pages}{217002} (\bibinfo{year}{2006}).

\bibitem[{\citenamefont{Lamura et~al.}(2006)\citenamefont{Lamura, Aurino,
  Cifariello, Di~Gennaro, Andreone, Emery, H\'erold, Mar\^ech\'e, and
  Lagrange}}]{lamura2006}
\bibinfo{author}{\bibfnamefont{G.}~\bibnamefont{Lamura}},
  \bibinfo{author}{\bibfnamefont{M.}~\bibnamefont{Aurino}},
  \bibinfo{author}{\bibfnamefont{G.}~\bibnamefont{Cifariello}},
  \bibinfo{author}{\bibfnamefont{E.}~\bibnamefont{Di~Gennaro}},
  \bibinfo{author}{\bibfnamefont{A.}~\bibnamefont{Andreone}},
  \bibinfo{author}{\bibfnamefont{N.}~\bibnamefont{Emery}},
  \bibinfo{author}{\bibfnamefont{C.}~\bibnamefont{H\'erold}},
  \bibinfo{author}{\bibfnamefont{J.-F.} \bibnamefont{Mar\^ech\'e}},
  \bibnamefont{and} \bibinfo{author}{\bibfnamefont{P.}~\bibnamefont{Lagrange}},
  \bibinfo{journal}{Phys. Rev. Lett.} \textbf{\bibinfo{volume}{96}},
  \bibinfo{pages}{107008} (\bibinfo{year}{2006}).

\bibitem[{\citenamefont{Hinks et~al.}()\citenamefont{Hinks, Rosenmann, Claus,
  Bailey, and Jorgensen}}]{hinks2007}
\bibinfo{author}{\bibfnamefont{D.~G.} \bibnamefont{Hinks}},
  \bibinfo{author}{\bibfnamefont{D.}~\bibnamefont{Rosenmann}},
  \bibinfo{author}{\bibfnamefont{H.}~\bibnamefont{Claus}},
  \bibinfo{author}{\bibfnamefont{M.~S.} \bibnamefont{Bailey}},
  \bibnamefont{and} \bibinfo{author}{\bibfnamefont{J.~D.}
  \bibnamefont{Jorgensen}} (????).

\bibitem[{\citenamefont{Mazin}(2005)}]{mazin2005}
\bibinfo{author}{\bibfnamefont{I.~I.} \bibnamefont{Mazin}},
  \bibinfo{journal}{Phys. Rev. Lett.} \textbf{\bibinfo{volume}{95}},
  \bibinfo{pages}{227001} (\bibinfo{year}{2005}).

\bibitem[{\citenamefont{Calandra and Mauri}(2005)}]{calandra2005}
\bibinfo{author}{\bibfnamefont{M.}~\bibnamefont{Calandra}} \bibnamefont{and}
  \bibinfo{author}{\bibfnamefont{F.}~\bibnamefont{Mauri}},
  \bibinfo{journal}{Phys. Rev. Lett.} \textbf{\bibinfo{volume}{95}},
  \bibinfo{pages}{237002} (\bibinfo{year}{2005}).

\bibitem[{\citenamefont{Boeri et~al.}(2007)\citenamefont{Boeri, Bachelet,
  Giantomassi, and Andersen}}]{boeri2007}
\bibinfo{author}{\bibfnamefont{L.}~\bibnamefont{Boeri}},
  \bibinfo{author}{\bibfnamefont{G.~B.} \bibnamefont{Bachelet}},
  \bibinfo{author}{\bibfnamefont{M.}~\bibnamefont{Giantomassi}},
  \bibnamefont{and} \bibinfo{author}{\bibfnamefont{O.~K.}
  \bibnamefont{Andersen}}, \bibinfo{journal}{Phys. Rev. B}
  \textbf{\bibinfo{volume}{76}}, \bibinfo{pages}{064510}
  (\bibinfo{year}{2007}).

\bibitem[{\citenamefont{Mazin et~al.}(2007)\citenamefont{Mazin, Boeri, Dolgov,
  Golubov, Bachelet, Giantomassi, and Andersen}}]{mazin2007}
\bibinfo{author}{\bibfnamefont{I.~I.} \bibnamefont{Mazin}},
  \bibinfo{author}{\bibfnamefont{L.}~\bibnamefont{Boeri}},
  \bibinfo{author}{\bibfnamefont{O.~V.} \bibnamefont{Dolgov}},
  \bibinfo{author}{\bibfnamefont{A.~A.} \bibnamefont{Golubov}},
  \bibinfo{author}{\bibfnamefont{G.~B.} \bibnamefont{Bachelet}},
  \bibinfo{author}{\bibfnamefont{M.}~\bibnamefont{Giantomassi}},
  \bibnamefont{and} \bibinfo{author}{\bibfnamefont{O.~K.}
  \bibnamefont{Andersen}}, \bibinfo{journal}{Physica C}
  \textbf{\bibinfo{volume}{460}}, \bibinfo{pages}{116} (\bibinfo{year}{2007}).

\bibitem[{\citenamefont{Posternak et~al.}(1983)\citenamefont{Posternak,
  Baldereschi, Freeman, Wimmer, and Weinert}}]{posternak1983}
\bibinfo{author}{\bibfnamefont{M.}~\bibnamefont{Posternak}},
  \bibinfo{author}{\bibfnamefont{A.}~\bibnamefont{Baldereschi}},
  \bibinfo{author}{\bibfnamefont{A.~J.} \bibnamefont{Freeman}},
  \bibinfo{author}{\bibfnamefont{E.}~\bibnamefont{Wimmer}}, \bibnamefont{and}
  \bibinfo{author}{\bibfnamefont{M.}~\bibnamefont{Weinert}},
  \bibinfo{journal}{Phys. Rev. Lett.} \textbf{\bibinfo{volume}{50}},
  \bibinfo{pages}{761} (\bibinfo{year}{1983}).

\bibitem[{\citenamefont{Holzwarth et~al.}(1984)\citenamefont{Holzwarth, Louie,
  and Rabii}}]{holzwarth1984}
\bibinfo{author}{\bibfnamefont{N.~A.~W.} \bibnamefont{Holzwarth}},
  \bibinfo{author}{\bibfnamefont{S.~G.} \bibnamefont{Louie}}, \bibnamefont{and}
  \bibinfo{author}{\bibfnamefont{S.}~\bibnamefont{Rabii}},
  \bibinfo{journal}{Phys. Rev. B} \textbf{\bibinfo{volume}{30}},
  \bibinfo{pages}{2219} (\bibinfo{year}{1984}).

\bibitem[{\citenamefont{Csanyi et~al.}(2005)\citenamefont{Csanyi, Littlewood,
  Nevidomskyy, Pickard, and Simons}}]{csanyi2005}
\bibinfo{author}{\bibfnamefont{G.}~\bibnamefont{Csanyi}},
  \bibinfo{author}{\bibfnamefont{P.}~\bibnamefont{Littlewood}},
  \bibinfo{author}{\bibfnamefont{A.}~\bibnamefont{Nevidomskyy}},
  \bibinfo{author}{\bibfnamefont{C.}~\bibnamefont{Pickard}}, \bibnamefont{and}
  \bibinfo{author}{\bibfnamefont{B.}~\bibnamefont{Simons}},
  \bibinfo{journal}{Nat. Phys.} \textbf{\bibinfo{volume}{1}},
  \bibinfo{pages}{42} (\bibinfo{year}{2005}).

\bibitem[{\citenamefont{Valla et~al.}(2009)\citenamefont{Valla, Camacho, Pan,
  Fedorov, Walters, Howard, and Ellerby}}]{valla2009}
\bibinfo{author}{\bibfnamefont{T.}~\bibnamefont{Valla}},
  \bibinfo{author}{\bibfnamefont{J.}~\bibnamefont{Camacho}},
  \bibinfo{author}{\bibfnamefont{Z.-H.} \bibnamefont{Pan}},
  \bibinfo{author}{\bibfnamefont{A.~V.} \bibnamefont{Fedorov}},
  \bibinfo{author}{\bibfnamefont{A.~C.} \bibnamefont{Walters}},
  \bibinfo{author}{\bibfnamefont{C.~A.} \bibnamefont{Howard}},
  \bibnamefont{and} \bibinfo{author}{\bibfnamefont{M.}~\bibnamefont{Ellerby}},
  \bibinfo{journal}{Phys. Rev. Lett.} \textbf{\bibinfo{volume}{102}},
  \bibinfo{pages}{107007} (\bibinfo{year}{2009}).

\bibitem[{\citenamefont{Gr\"uneis et~al.}(2009)\citenamefont{Gr\"uneis,
  Attaccalite, Rubio, Vyalikh, Molodtsov, Fink, Follath, Eberhardt, B\"uchner,
  and Pichler}}]{gruneis2009}
\bibinfo{author}{\bibfnamefont{A.}~\bibnamefont{Gr\"uneis}},
  \bibinfo{author}{\bibfnamefont{C.}~\bibnamefont{Attaccalite}},
  \bibinfo{author}{\bibfnamefont{A.}~\bibnamefont{Rubio}},
  \bibinfo{author}{\bibfnamefont{D.~V.} \bibnamefont{Vyalikh}},
  \bibinfo{author}{\bibfnamefont{S.~L.} \bibnamefont{Molodtsov}},
  \bibinfo{author}{\bibfnamefont{J.}~\bibnamefont{Fink}},
  \bibinfo{author}{\bibfnamefont{R.}~\bibnamefont{Follath}},
  \bibinfo{author}{\bibfnamefont{W.}~\bibnamefont{Eberhardt}},
  \bibinfo{author}{\bibfnamefont{B.}~\bibnamefont{B\"uchner}},
  \bibnamefont{and} \bibinfo{author}{\bibfnamefont{T.}~\bibnamefont{Pichler}},
  \bibinfo{journal}{Phys. Rev. B} \textbf{\bibinfo{volume}{80}},
  \bibinfo{pages}{075431} (\bibinfo{year}{2009}).

\bibitem[{\citenamefont{Dean et~al.}(2010)\citenamefont{Dean, Howard, Saxena,
  and Ellerby}}]{dean2010}
\bibinfo{author}{\bibfnamefont{M.~P.~M.} \bibnamefont{Dean}},
  \bibinfo{author}{\bibfnamefont{C.~A.} \bibnamefont{Howard}},
  \bibinfo{author}{\bibfnamefont{S.~S.} \bibnamefont{Saxena}},
  \bibnamefont{and} \bibinfo{author}{\bibfnamefont{M.}~\bibnamefont{Ellerby}},
  \bibinfo{journal}{Phys. Rev. B} \textbf{\bibinfo{volume}{81}},
  \bibinfo{pages}{045405} (\bibinfo{year}{2010}).

\bibitem[{\citenamefont{Pruvost et~al.}(2004)\citenamefont{Pruvost, Herold,
  Herold, and Lagrange}}]{pruvost2004}
\bibinfo{author}{\bibfnamefont{S.}~\bibnamefont{Pruvost}},
  \bibinfo{author}{\bibfnamefont{C.}~\bibnamefont{Herold}},
  \bibinfo{author}{\bibfnamefont{A.}~\bibnamefont{Herold}}, \bibnamefont{and}
  \bibinfo{author}{\bibfnamefont{P.}~\bibnamefont{Lagrange}},
  \bibinfo{journal}{Carbon} \textbf{\bibinfo{volume}{42}},
  \bibinfo{pages}{1825} (\bibinfo{year}{2004}).

\bibitem[{\citenamefont{Sugawara et~al.}(2009)\citenamefont{Sugawara, Sato, and
  Takahashi}}]{sugawara2009}
\bibinfo{author}{\bibfnamefont{K.}~\bibnamefont{Sugawara}},
  \bibinfo{author}{\bibfnamefont{T.}~\bibnamefont{Sato}}, \bibnamefont{and}
  \bibinfo{author}{\bibfnamefont{T.}~\bibnamefont{Takahashi}},
  \bibinfo{journal}{Nat. Phys.} \textbf{\bibinfo{volume}{5}},
  \bibinfo{pages}{40} (\bibinfo{year}{2009}).

\bibitem[{\citenamefont{Valla et~al.}(1999)\citenamefont{Valla, Fedorov,
  Johnson, and Hulbert}}]{valla1999}
\bibinfo{author}{\bibfnamefont{T.}~\bibnamefont{Valla}},
  \bibinfo{author}{\bibfnamefont{A.~V.} \bibnamefont{Fedorov}},
  \bibinfo{author}{\bibfnamefont{P.~D.} \bibnamefont{Johnson}},
  \bibnamefont{and} \bibinfo{author}{\bibfnamefont{S.~L.}
  \bibnamefont{Hulbert}}, \bibinfo{journal}{Phys. Rev. Lett.}
  \textbf{\bibinfo{volume}{83}}, \bibinfo{pages}{2085} (\bibinfo{year}{1999}).

\bibitem[{\citenamefont{LaShell et~al.}(2000)\citenamefont{LaShell, Jensen, and
  Balasubramanian}}]{lashell2000}
\bibinfo{author}{\bibfnamefont{S.}~\bibnamefont{LaShell}},
  \bibinfo{author}{\bibfnamefont{E.}~\bibnamefont{Jensen}}, \bibnamefont{and}
  \bibinfo{author}{\bibfnamefont{T.}~\bibnamefont{Balasubramanian}},
  \bibinfo{journal}{Phys. Rev. B} \textbf{\bibinfo{volume}{61}},
  \bibinfo{pages}{2371} (\bibinfo{year}{2000}).

\bibitem[{\citenamefont{Valla et~al.}(2000)\citenamefont{Valla, Fedorov,
  Johnson, Li, Gu, and Koshizuka}}]{valla2000}
\bibinfo{author}{\bibfnamefont{T.}~\bibnamefont{Valla}},
  \bibinfo{author}{\bibfnamefont{A.~V.} \bibnamefont{Fedorov}},
  \bibinfo{author}{\bibfnamefont{P.~D.} \bibnamefont{Johnson}},
  \bibinfo{author}{\bibfnamefont{Q.}~\bibnamefont{Li}},
  \bibinfo{author}{\bibfnamefont{G.~D.} \bibnamefont{Gu}}, \bibnamefont{and}
  \bibinfo{author}{\bibfnamefont{N.}~\bibnamefont{Koshizuka}},
  \bibinfo{journal}{Phys. Rev. Lett.} \textbf{\bibinfo{volume}{85}},
  \bibinfo{pages}{828} (\bibinfo{year}{2000}).

\bibitem[{\citenamefont{Kordyuk et~al.}(2005)\citenamefont{Kordyuk, Borisenko,
  Koitzsch, Fink, Knupfer, and Berger}}]{kordyuk2005}
\bibinfo{author}{\bibfnamefont{A.~A.} \bibnamefont{Kordyuk}},
  \bibinfo{author}{\bibfnamefont{S.~V.} \bibnamefont{Borisenko}},
  \bibinfo{author}{\bibfnamefont{A.}~\bibnamefont{Koitzsch}},
  \bibinfo{author}{\bibfnamefont{J.}~\bibnamefont{Fink}},
  \bibinfo{author}{\bibfnamefont{M.}~\bibnamefont{Knupfer}}, \bibnamefont{and}
  \bibinfo{author}{\bibfnamefont{H.}~\bibnamefont{Berger}},
  \bibinfo{journal}{Phys. Rev. B} \textbf{\bibinfo{volume}{71}},
  \bibinfo{pages}{214513} (\bibinfo{year}{2005}).

\bibitem[{\citenamefont{McMillan}(1968)}]{McMillan1968}
\bibinfo{author}{\bibfnamefont{W.}~\bibnamefont{McMillan}},
  \bibinfo{journal}{Phys. Rev.} \textbf{\bibinfo{volume}{167}},
  \bibinfo{pages}{331} (\bibinfo{year}{1968}).

\bibitem[{\citenamefont{Camacho et~al.}()\citenamefont{Camacho, Upton, Pan,
  Walters, Howard, Ellerby, and Valla}}]{camacho}
\bibinfo{author}{\bibfnamefont{J.}~\bibnamefont{Camacho}},
  \bibinfo{author}{\bibfnamefont{M.}~\bibnamefont{Upton}},
  \bibinfo{author}{\bibfnamefont{Z.-H.} \bibnamefont{Pan}},
  \bibinfo{author}{\bibfnamefont{A.}~\bibnamefont{Walters}},
  \bibinfo{author}{\bibfnamefont{C.}~\bibnamefont{Howard}},
  \bibinfo{author}{\bibfnamefont{M.}~\bibnamefont{Ellerby}}, \bibnamefont{and}
  \bibinfo{author}{\bibfnamefont{T.}~\bibnamefont{Valla}},
  \bibinfo{note}{unpublished}.

\bibitem[{\citenamefont{Valla}(2006)}]{valla2006}
\bibinfo{author}{\bibfnamefont{T.}~\bibnamefont{Valla}},
  \bibinfo{journal}{Phys. Rev. Lett.} \textbf{\bibinfo{volume}{96}},
  \bibinfo{pages}{119701} (\bibinfo{year}{2006}).

\bibitem[{\citenamefont{McChesney et~al.}(2007)\citenamefont{McChesney,
  Bostwick, Ohta, Emtsev, Seyller, Horn, and Rotenberg}}]{mcchesney2007}
\bibinfo{author}{\bibfnamefont{J.~L.} \bibnamefont{McChesney}},
  \bibinfo{author}{\bibfnamefont{A.}~\bibnamefont{Bostwick}},
  \bibinfo{author}{\bibfnamefont{T.}~\bibnamefont{Ohta}},
  \bibinfo{author}{\bibfnamefont{K.~V.} \bibnamefont{Emtsev}},
  \bibinfo{author}{\bibfnamefont{T.}~\bibnamefont{Seyller}},
  \bibinfo{author}{\bibfnamefont{K.}~\bibnamefont{Horn}}, \bibnamefont{and}
  \bibinfo{author}{\bibfnamefont{E.}~\bibnamefont{Rotenberg}},
  \bibinfo{journal}{arXiv:0705.3264}  (\bibinfo{year}{2007}).

\bibitem[{\citenamefont{Leem et~al.}(2008)\citenamefont{Leem, Kim, Kim, Park,
  Ohta, Bostwick, Rotenberg, Kim, Kim, Choi et~al.}}]{leem2008}
\bibinfo{author}{\bibfnamefont{C.~S.} \bibnamefont{Leem}},
  \bibinfo{author}{\bibfnamefont{B.~J.} \bibnamefont{Kim}},
  \bibinfo{author}{\bibfnamefont{C.}~\bibnamefont{Kim}},
  \bibinfo{author}{\bibfnamefont{S.~R.} \bibnamefont{Park}},
  \bibinfo{author}{\bibfnamefont{T.}~\bibnamefont{Ohta}},
  \bibinfo{author}{\bibfnamefont{A.}~\bibnamefont{Bostwick}},
  \bibinfo{author}{\bibfnamefont{E.}~\bibnamefont{Rotenberg}},
  \bibinfo{author}{\bibfnamefont{H.~D.} \bibnamefont{Kim}},
  \bibinfo{author}{\bibfnamefont{M.~K.} \bibnamefont{Kim}},
  \bibinfo{author}{\bibfnamefont{H.~J.} \bibnamefont{Choi}},
  \bibnamefont{et~al.}, \bibinfo{journal}{Phys. Rev. Lett.}
  \textbf{\bibinfo{volume}{100}}, \bibinfo{pages}{016802}
  (\bibinfo{year}{2008}).

\end{thebibliography}
\providecommand{\noopsort}[1]{}\providecommand{\singleletter}[1]{#1}%

\end{document}